\documentclass[a4paper]{raa}
\usepackage{graphicx,times}
\usepackage{natbib}
\usepackage{amssymb,amsmath}
\bibpunct{(}{)}{;}{a}{}{,}

\usepackage{hyperref}
\hypersetup{pdftex=true,pagebackref=true, colorlinks = true, linkcolor = green, anchorcolor = red, citecolor = blue, filecolor = red, pagecolor = red, urlcolor = red}

\begin{document}

  \title{NBFTP: A Dedicated Data Transfer System for Remote Astronomical Observation at Dome A $^*$
  \footnotetext{\small $*$ The source code has been upload
    on gitee (\url{https://gitee.com/AstroTJU/NBFTP}) and china-vo (\url{https://code.china-vo.org/ast3/nbftp}).}
    }

  \volnopage{ {\bf 20XX} Vol.\ {\bf X} No. {\bf XX}, 000--000}
  \setcounter{page}{1}

  \author{Siyuan Huang\inst{1}, Ce Yu\inst{1,2},
    Chao Sun$^{**}$\inst{1,2}, Yi Hu$^{**}$\inst{3},
    Zhaohui Shang\inst{3,4}, Bin Ma\inst{3},
    Ming Che\inst{1}, Xiaoxiao Lu\inst{1}
    \footnotetext{\small $**$ Corresponding author: Chao Sun (sch@tju.edu.cn), Yi Hu (huyi.naoc@gmail.com)}
  }

  \institute{ College of Intelligence and Computing,
    Tianjin University, 135 Yaguan Road Jinnan District,
    Tianjin 300350, China; {\it sch@tju.edu.cn}\\
    \and
    NAOC-TJU Joint Research Center in Astro-Informatic,
    135 Yaguan Road Jinnan District, Tianjin 300350, China\\
    \and
    National Astronomical Observatories,
    Chinese Academy of Science, 20 Datun Road Chaoyang District,
    Beijing 100012, China; {\it huyi.naoc@gmail.com}\\
    \and
    Tianjin Normal University, 393 Binshuixidao, Tianjin 300387, China\\
    \vs \no
    {\small Received 20XX Month Day; accepted 20XX Month Day}
  }

  \abstract{Dome A, Antarctica has been thought to be one of the best astronomical sites on the earth since decades ago. From it was first visited by astronomers in 2008, dozens of facilities for astronomical observation and site testing were deployed. Due to its special geographical location, the data and message exchange between Dome A and the domestic control center could only depend on Iridium. Because the link bandwidth of Iridium is extremely limited, meanwhile the network traffic cost is quite expensive and the network is rather unstable, the commonly used data transfer tools, such as rsync and scp, are not suitable in this case. In this paper, we design and implement a data transfer tool called NBFTP (narrow bandwidth file transfer protocol) for the astronomical observation of Dome A. NBFTP uses a uniform interface to arrange all types of data and matches specific transmission schemes for different data types according to rules. Break-point resuming and extensibility functions are also implemented. Our experimental results show that NBFTP consumes 60\% less network traffic than rsync when detecting the data pending to be transferred. And when transferring small files of 1KB, the network traffic consumption of NBFTP is 40\% less than rsync. However, as the file size increases, the network traffic consumption of NBFTP tends to approach rsync, but it is still smaller than rsync.
  }

  \authorrunning{S. Huang, C. Yu, C. Sun, Y. Hu, Z. Shang, B. Ma et al. }
  \titlerunning{NBFTP: A Dedicated Data Transfer System}
  \maketitle

\section{Introduction}
\label{sec:sec1}

Astronomical observation always needs the most strict requirements of the experimental environment. Modern astronomical sites are usually selected as the most isolated places on the earth, to get rid of the influence of human activities and obtain supreme observational conditions. For example, Dome A, the highest point on the Antarctic inland ice cap, is potentially one of the best astronomical observation sites on the planet (\citealt{Saunders+etal+2009, Hu+etal+2014, Hu+etal+2019}). With its extremely stable atmosphere and long polar nights, the site condition at Dome A could even be comparable to that in the space. The 21st Chinese National Antarctic Research Expedition (CHINARE) reached Dome A for the first time in 2005, and since the 24th CHINARE reached Dome A for the second time, dozens of facilities for astronomical observation and site testing were deployed (\citealt{Yuan+etal+2008, Yuan+etal+2010, Bonner+etal+2010, Hu+etal+2014, Shi+etal+2016, Ma+etal+2018, Shang+etal+2018}).

However, because of the harsh environment and logistical difficulties, Dome A is still a completely unattended site, which means that there will be no human on-site to operate any facilities after the expedition team leaves at the end of every January (\citealt{Hu+etal+2019}). And this situation will still be true for the foreseeable future. Although the facilities installed at Dome A were designed to operate as automatically as possible (\citealt{Hu+etal+2016}), it is still inevitable to transmit data back and forth between Dome A and the domestic control center (DCC). Thus, an effective method of data transfer is quite crucial for safely operating the remote astronomical observatory at Dome A (\citealt{Zhang+etal+2016, Kubanek+2016}).

As a fast and versatile incremental file transfer tool, rsync is probably the best choice for data transfer under various network conditions (\citealt{Shial+2015}). rsync is known for its delta transfer algorithm. It firstly compares the files in both source and target, and then only sends the differences between them so that it reduces the amount of data to transfer\footnote{\url{https://linux.die.net/man/1/rsync}}. However, this tool does not achieve network traffic economization in all the cases, because it does not truly implement the network traffic control. When transferring small files, rsync performs unsatisfactorily due to the overhead of the file comparison and incremental calculations. It will also consume extra network traffic when checking the differences directories between the source and target.

Located in the innermost of the continent, Dome A has no other Internet access except Iridium (\citealt{Lawrence+etal+2008, Jia+etal+2018}). The power and Iridium connection for instruments are provided by an automated observatory platform PLATO-A (\citealt{Ashley+etal+2010}). Because the Iridium network communication is not only unstable but also very expensive (\citealt{Shang+etal+2012}), maximizing the utility of the Iridium channel is quite important for us to operate our observatory at Dome A. Most of the data transferred from Dome A to the DCC are system log files, stamp images, site-testing data, and alarm messages (\citealt{Hu+etal+2016}). The typical size of these data is around 10KB. As we showed above, rsync will use the Iridium channel inefficiently because of its severe overhead. Hence, it is urgent to develop a new tool for transferring data, especially for files with small size under Iridium network communication.

Therefore, we design NBFTP, which is an intelligent data transfer system that achieves fine-grained network traffic control. The NBFTP system is composed of a sender at Dome A and a receiver at the DCC. It can effectively transmit the data from the observatory at Dome A to the DCC with user-provided priority.

The rest of this article is organized as follows. We show the related works on data transfer in Section \ref{sec:sec2}. We then describe the architecture of NBFTP and its detailed information on modules and algorithms in Section \ref{sec:sec3}.  And we show the performance of NBFTP and compare NBFTP with other file transfer tools in Section \ref{sec:sec4}. In the final section, we summarize and discuss future work.

\section{Related Work}
\label{sec:sec2}

Data transfer is one of the basic functions of computer networks. There are mature file transfer protocols such as File Transfer Protocol (FTP) and SSH File Transfer Protocol (SFTP), and most Linux distributions have integrated practical tools, such as rcp, scp, rsync, etc.

FTP is a standard network protocol for transfer files between clients and servers on a computer network\footnote{\url{https://en.wikipedia.org/wiki/File_Transfer_Protocol}}. The main advantage of FTP is that it comes with the most commonly used operating systems, including most Linux distributions and Microsoft Windows. Most FTP-based tools support resuming from break-points. SFTP is a component of Secure Shell (SSH), which supports encrypted file transfer and resuming\footnote{\url{https://en.wikipedia.org/wiki/SSH_File_Transfer_Protocol}}. The encryption and decryption technology of SFTP makes it more secure than FTP, but the process also creates another kind of overhead. Besides, file transfer based on FTP or SFTP requires interactive operations even when used in automated scripts. These  interaction operations will cause extra network traffic.

As a remote file copy tool, rcp can transfer files or directories between different hosts\footnote{\url{https://en.wikipedia.org/wiki/Berkeley_r-commands}}. Scp encrypted with SSH is a more secure version of rcp\footnote{\url{https://en.wikipedia.org/wiki/Secure_copy}}. Both tools copy all the files in the source directory, including those have been transferred. Thus, they may introduce unnecessary traffic consumption. Besides, neither tools support break-point resuming.

Rsync is designed to synchronize file directories between different hosts\footnote{\url{https://linux.die.net/man/1/rsync}}. Its incremental transfer algorithm reduces the amount of data transmitted over the network by only sending the different parts between the source files and the existing files in the destination. This approach makes it consume less network traffic than which rcp and scp do. Unlike rcp and scp, rsync supports break-point resuming. However, the incremental algorithm of rsync is not suitable for the situation where the number of files keeps increasing. Every time rsync is called, it will check all subdirectories and files in the directories at both ends, causing unnecessary overhead. rsync supports resuming by storing extra information in a temporary file. Therefore, it needs to perform additional file operations and information calculations every time it transfers a file. Such a resuming procedure may decrease the transfer efficiency.

\section{TRAFFIC-SAVING DATA TRANSFER SYSTEM}
\label{sec:sec3}

\subsection{NBFTP Software Architecture}

\begin{figure}
   \centering
   \includegraphics[width=\linewidth]{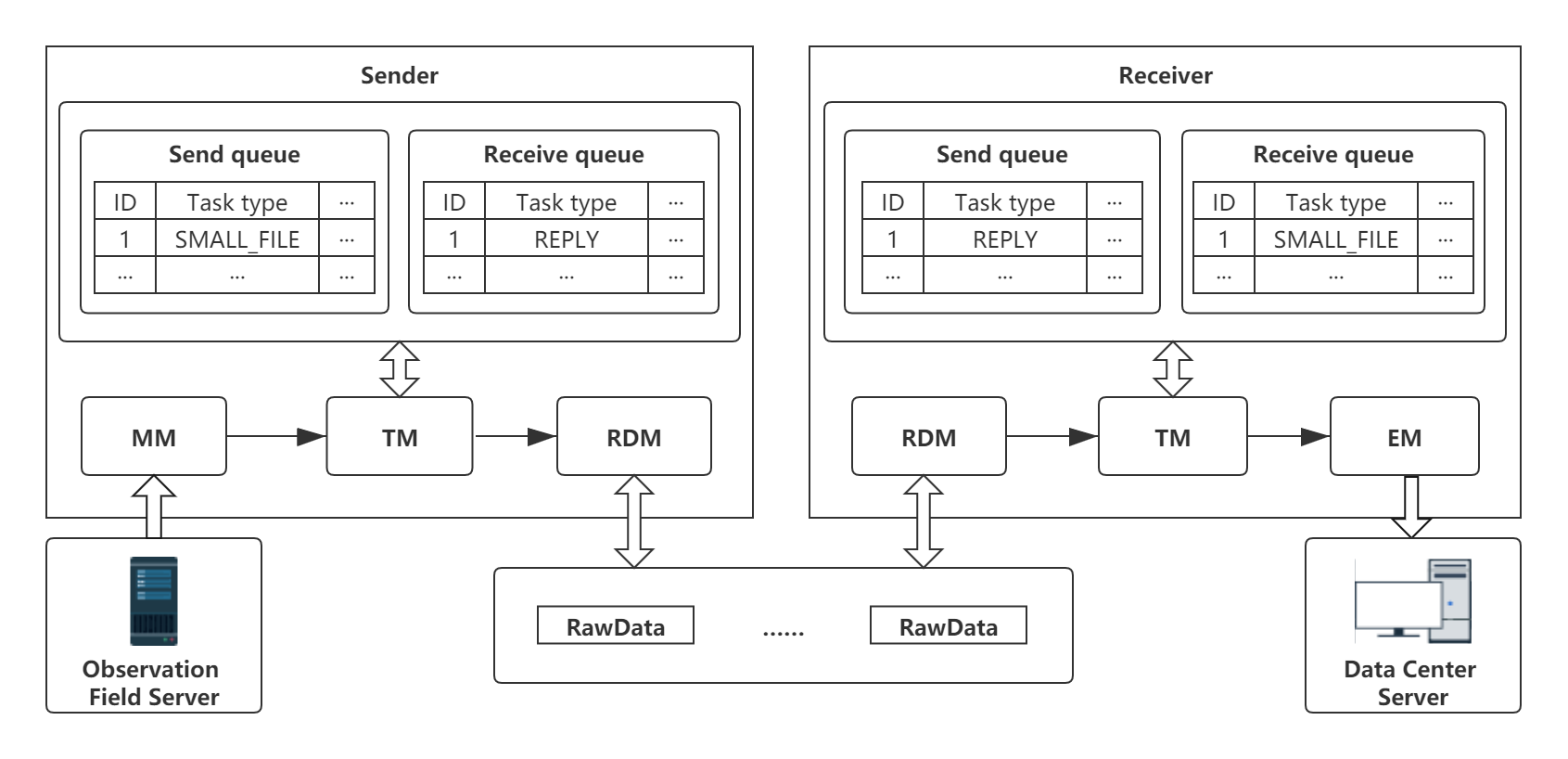}
   \caption{The software architecture of NBFTP.}
   \label{fig:fig1}
\end{figure}

Our work focuses on transferring data from Dome A to the DCC stably and cost-effectively. As shown in Figure~\ref{fig:fig1}, NBFTP consists of a sender and a receiver. The sender of NBFTP is deployed at the Dome A. It includes a Monitor Module (MM), a Task Module (TM), and a Data Module (DM). The receiver is deployed at the DCC, which mainly includes a TM, an DM, and an Expansion Module (EM). It is loosely-coupled between the components which can effectively reduce the scale and complexity of the system. Although it is the sender that transferring data to the receiver, the receiver can also transfer files to the sender.

Once the MM detects a file to transfer, it will add a transmission task to the task queue. The TM is responsible for managing the task queue, such as sending and receiving data, sorting the tasks in the queues according to priorities. We will interpret task types in Section \ref{sec:sec3.2}. The DM provides a series of functions for packing data to or extracting from the buffer. The EM is used by the receiver to complete other functions using shell commands after receiving a file or message, such as sending mail and importing the received data to a database.

Overall, the main functions of NBFTP include:
\begin{itemize}
  \item Optimizing data size / actual network traffic. In the TM, NBFTP intelligently matches the throttling transmission scheme according to the size of the data to be transmitted and assigns transmission tasks.
  \item Resuming break-points. In the TM, all tasks will be tracked to the block level. Only after each block is successfully transmitted can the next block be transferred. So that even after disconnection, the last position of the successful transmission can be found to start again.
  \item Extensibility functions. In the EM, we can start the automatic transaction processing, such as sending an email, by modifying the configuration file of the domestic receiver.
\end{itemize}

NBFTP is a tool for transferring data between Dome A and the DCC. It is an open-source project running on the Linux platform, coded in C++, with the following dependencies: GNU C Library, GNU GLIB 2.0, C Shell, Gamin. There are two programs of nbftp and nbftp-server in our project. nbftp-server is a daemon process, which starts immediately after the operating system is booted up. Given different configuration parameters, nbftp-server can act as either the sender or the receiver. Users can execute nbftp commands, which is the client of nbftp-server, instead of directly manipulating the file directory. The operations of sending data and receiving data are different. The specific instructions and their meanings are tabulated in Table~\ref{tab:tab1} and ~\ref{tab:tab2}.

\begin{table}
\bc
\begin{minipage}[]{\linewidth}
\caption[]{Operations of sending data\label{tab:tab1}}\end{minipage}
\setlength{\tabcolsep}{5pt}
\small
  \begin{tabular}{lll}
    \hline\noalign{\smallskip}
    Instruction & Meaning & Remark \\
    \hline\noalign{\smallskip}
    nbftp push FILENAME [PRIORITY] & Submit a file to transfer & The default priority is 5 \\
    nbftp mail MASSAGE & Submit a massage to transfer & The default priority is 3 \\
    nbftp pull TASKID & Cancel file transfer task & \\
    nbftp pull all & Cancel all transfer tasks & \\
    nbftp pending & List pending files & \\
    \noalign{\smallskip}\hline
  \end{tabular}
\ec
\end{table}

\begin{table}
\bc
\begin{minipage}[]{\linewidth}
\caption[]{Operations of receiving data\label{tab:tab2}}\end{minipage}
\setlength{\tabcolsep}{5pt}
\small
  \begin{tabular}{lll}
    \hline\noalign{\smallskip}
    Instruction & Meaning & Remark \\
    \hline\noalign{\smallskip}
    nbftp list & List currently received files & \\
    nbftp get TASKID & Save the file to the current directory or show massage & \\
    nbftp delete TASKID & Delete the reception record & \\
    nbftp delete all & Delete all reception records & \\
    \noalign{\smallskip}\hline
  \end{tabular}
\ec
\end{table}

\subsection{Transfer Protocol}
\label{sec:sec3.2}

The TM of NBFTP is carefully designed to use unified interfaces to arrange all types of data, corresponding to eight task types: SMALL FILE, LARGE FILE, MESSAGE, LASTPOS, COMPLETE SIG, CHECK PASS, CHECK FAILED and REPLY. The first three types are used to transfer small files, large files, and messages, while the latter five are used for transmitting feedback information to ensure the stability of the transmission. During a transmission process, each task of the sender will be packaged into one or more RawData segments, which are the smallest units of transmission. Then the receiver will resemble it. Different tasks follow different packing rules. The communication protocol of this transmission system is tabulated in Table~\ref{tab:tab3}.

\begin{table}
\bc
\begin{minipage}[]{\linewidth}
\caption[]{NBFTP transfer protocol\label{tab:tab3}}\end{minipage}
\setlength{\tabcolsep}{5pt}
\small
  \begin{tabular}{ccl}
    \hline\noalign{\smallskip}
    RawData type & Data length (Byte) & Content \\
    \hline\noalign{\smallskip}
      & 1 & sign: SMALL FILE DATA \\
      & 4 & task id \\
    SMALL FILE & 16 & MD5 \\
      & 1 + name\_length & file name \\
      & 4 + file\_size & binary file content \\
    \hline
      & 1 & sign: LARGE FILE BEGIN \\
      & 4 & task ID \\
    LARGE FILE (begin) & 16 & MD5 \\
      & 1 + name\_length & file name \\
      & 4 & file size \\
    \hline
      & 1 & sign: LARGE FILE PART \\
    LARGE FILE (part) & 4 & task ID \\
      & 4 & current position \\
      & 4 + file\_block\_size & binary file block content \\
    \hline
      & 1 & sign: MESSAGE DATA \\
    MESSAGE & 4 & task ID \\
      & 1 + message\_length & message \\
    \hline
     & 1 & sign: LASTPOS / COMPLETE SIG / CHECK PASS /  \\
    LASTPOS / COMPLETE SIG / &  & CHECK FAILED / REPLY \\
    CHECK PASS / CHECK FAILED / & 4 & task ID \\
    REPLY & 4 & associated task ID \\
     & 4 & current position (LASTPOS only) \\
    \noalign{\smallskip}\hline
  \end{tabular}
\ec
\end{table}

For all the types of tasks, the first byte in NBFTP stores the type name, and the following four bytes store the ID of the task. The subsequent information will be different and correspond one-to-one in Table~\ref{tab:tab3}. For instance, when transferring a small size file, the TM of the sender uses only one RawData segment. It firstly sets the task type as SMALL FILE in the first byte and assigns a task ID. Then the TM calculates the MD5 checksum of the file and fills the value into the 16-byte MD5 field of the segment. Finally, it fills the length of the file name, the file name, the size of the file, and the file content into the segment. After these steps, the TM sends the RawData segment to the receiver. The TM of the receiver receives and extracts the segment, writes the content to a new file. Then it calculates the MD5 of the file and compares it with that in the MD5 field. If the file check is successful, the TM acknowledges a CHECK PASS segment to the sender to signify the success of transmission. Otherwise, it acknowledges a CHECK FAILED segment to the sender to ask for retransmitting.

It is worth mentioning that sending LARGE FILE tasks uses two types of RawData. The former is the LARGE FILE (begin), which contains metadata for large files, marking the beginning position of a large file chunked transfer task. Large files are divided into blocks, and each block uses the latter type of RawData named LARGE FILE (part), which is then transferred in turn. If the receiver TM receives a LARGE FILE (begin) or LARGE FILE (part) other than the last segment, it acknowledges a REPLY segment to the sender.

The five kinds of feedback information are sent by the receiver to indicate the current data reception status to the sender. Each of them uses 4 bytes to store the associated task ID. In particular, the LASTPOS uses 4 bytes to show the current transmission position, which can help to implement the break-point resuming function.

\subsection{Break-point Resuming Function}
\label{sec:sec3.3}

Since the Iridium network communication is extremely unstable, the packet loss must happen frequently. Therefore, any file transfer tools without supporting the break-point resuming function are not suitable for using at Dome A.

NBFTP implements break-point resuming function by dividing files into blocks. It then encapsulates the blocks into RawData segments, as tabulated in Table \ref{tab:tab3}. RawData segment is the smallest unit to transmit in NBFTP, which means it will be either delivered successfully or discarded completely. As we demonstrated in Section \ref{sec:sec3.2}, the sender of NBFTP issues only a single RawData segment for a small size file. On the contrary, for a large size file, the sender will firstly issue a LARGE FILE (begin) segment which contains only the metadata of that file; it then delivers the file body of the file by issuing LARGE FILE (part) segments. Each LARGE FILE (part) contains a ``current position'' field to record the position of the data block of the file. Whenever the sender delivers a segment, it will wait for a REPLY acknowledgment from the receiver. If the delivery fails or waiting for REPLY acknowledgment is timeout, the sender will suspend for a while and try to send the LARGE FILE (part) segment in the next transmission. In this case, the receiver will acknowledge a LASTPOS RawData segment rather than a REPLY segment to the sender. This segment also contains a ``current position'' field to signify the sender the restart to transfer the file from that position. By applying the LASTPOS segment, NBFTP implements break-point resuming.

Obviously, the size of the block will significantly influence the transmission efficiency of NBFTP. When the loss rate is high in a bad network condition, adopting a large block size will more likely waste the cost of Iridium network communication. On the other hand, using a small block size will introducing extra protocol overhead when encapsulating the RawData segments. Since this trade-off is strongly dependent on the network condition, we use an adaptive block size mechanism inspired by TCP congestion control\footnote{http://www.hjp.at/doc/rfc/rfc5681.html}. The block size is set to $2^{13}$ bytes at the beginning. After the previous block is successfully transmitted, which means that the network may be in good condition, then the block size will be doubled, reducing the block prefix consumption. If it reaches a maximum of $2^{22}$ bytes, it will not increase. If a block loss occurs midway, it means that the network environment becomes poor and the packet loss rate is increased, then the block size will shrink to reduce extra network traffic. The block size will be adjusted according to the network conditions, which is more suitable for unstable network environments. By dynamical determining the block size, NBFTP achieves a fine controlling network traffic.

\subsection{Extensibility Function}

In addition to transferring data with low network traffic consumption, NBFTP also provides an interface for users to implement extensibility functions, which are mainly used for subsequent processing after receiving data. After accepting a certain type of data, the receiver can automatically call other programs according to the configuration file, including but not limited to the following functions: proofreading data correctness, sending SMS or email to the specified location, importing the data in the file to the database, etc.

The implementation details of the extensibility functions are shown in Table~\ref{tab:tab4}.

\begin{table}
\bc
\begin{minipage}[]{\linewidth}
\caption[]{Functions of the Expansion Module\label{tab:tab4}}\end{minipage}
\setlength{\tabcolsep}{5pt}
\small
  \begin{tabular}{cl}
    \hline\noalign{\smallskip}
    Basic features of the Expansion Module & Notes \\
    \hline\noalign{\smallskip}
    Prerequisite & File reception completed \\
    Configuration setting & Set the command with specific parameters in the configuration file \\
    Supported program languages & C/C++/Shell/Java/Python/Perl/... \\
    Supported function & Mail sending/Data cleaning/Data archiving/SMS/... \\
    \noalign{\smallskip}\hline
  \end{tabular}
\ec
\end{table}

\subsection{Autonomous Data Transmission}
In order to implement fully automatic data transmission, it is necessary to ensure that the system can continue to be stable under unmanned operation and poor network conditions to ensure the integrity of the data. At the same time, special events can be sent to the DCC as soon as possible, to achieve a certain degree of resilience.

NBFTP implements fully automatic data transmission through MM. The main functions are shown in Table~\ref{tab:tab5}.

\begin{table}
\bc
\begin{minipage}[]{\linewidth}
\caption[]{Functions of the Monitor Module\label{tab:tab5}}\end{minipage}
\setlength{\tabcolsep}{5pt}
\small
  \begin{tabular}{cl}
    \hline\noalign{\smallskip}
    Event & Response\\
    \hline\noalign{\smallskip}
    Generate new observations & Start the data transfer process \\
    Generate specific types of data & Set a specific action (e.g. send back to country with highest priority) \\
    Tasks often terminate abnormally & Restart or terminate the task as needed \\
    \noalign{\smallskip}\hline
  \end{tabular}
\ec
\end{table}

\section{EXPERIMENTS AND RESULTS}
\label{sec:sec4}

\subsection{Experimental Settings}

Astronomical observations at Dome A require NBFTP to support fully automatic continuous operation for a whole year. According to the actual needs of data transmission, we analyzed and evaluated the performance of NBFTP in the following aspects: single file transfer, periodic file transfer, directory synchronization overhead, and file block parameter settings.

Most of the files transferred from Dome A are system log, site-testing data, and alarm messages. The system log and site-testing data will be transferred periodically. The compressed file size of the system log and site-testing data is about 8KB, and the size of the alarm message is less than 1KB. The transmission of observation data will be carried out as needed, and the transmission frequency and file size cannot be determined in advance. In order to simulate Iridium network communication, we limit the network bandwidth to 128Kbps.

As the main competitor, rsync is set to run in daemon mode which requires less network transmission than the interactive shell mode.

\subsection{Single File Transfer}

\begin{figure}[tbp]
	\centering
  \begin{minipage}{0.48\linewidth}
    \centering
    \includegraphics[width=\linewidth]{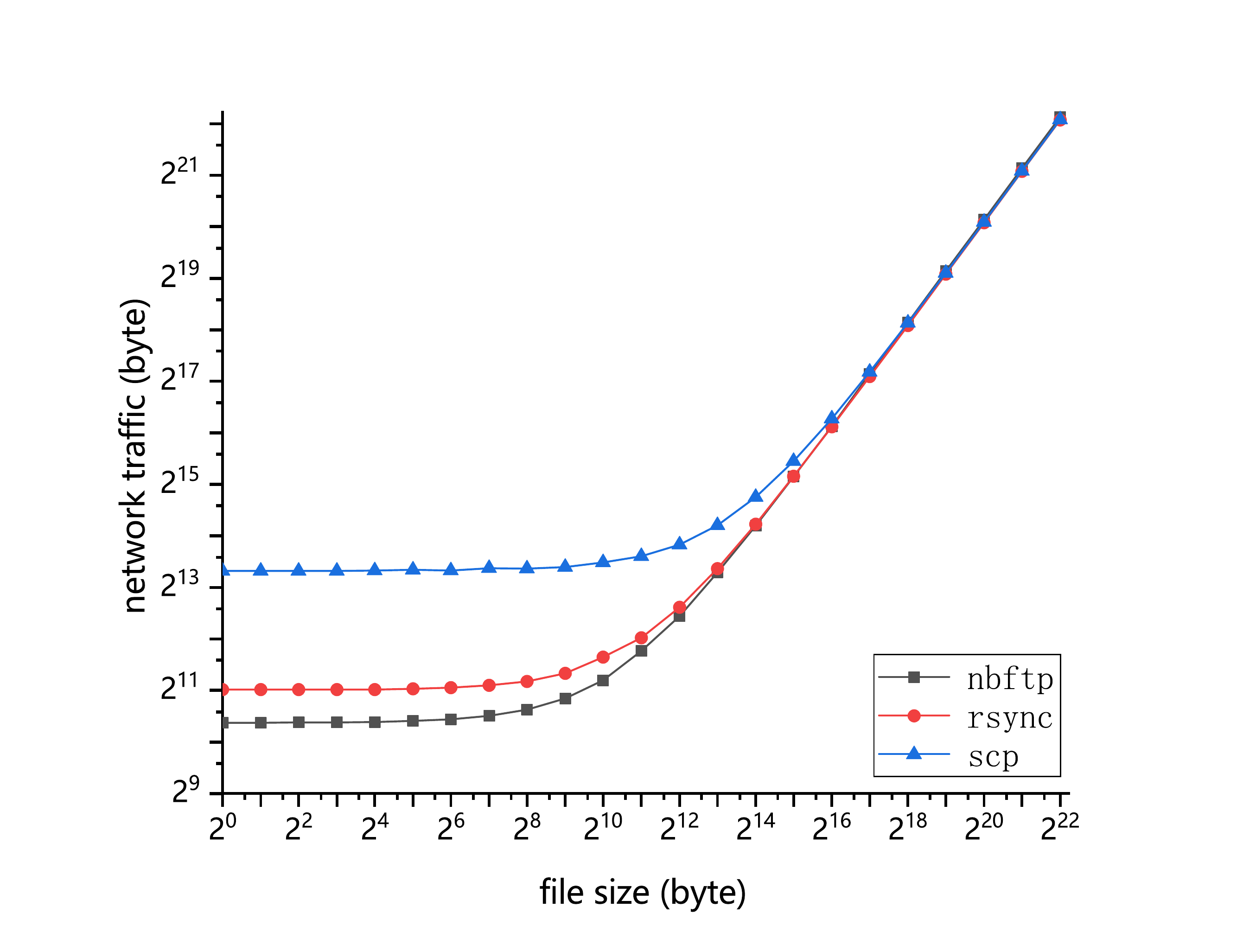}
    \subcaption{network traffic}
  \end{minipage}
  \begin{minipage}{0.48\linewidth}
    \centering
    \includegraphics[width=\linewidth]{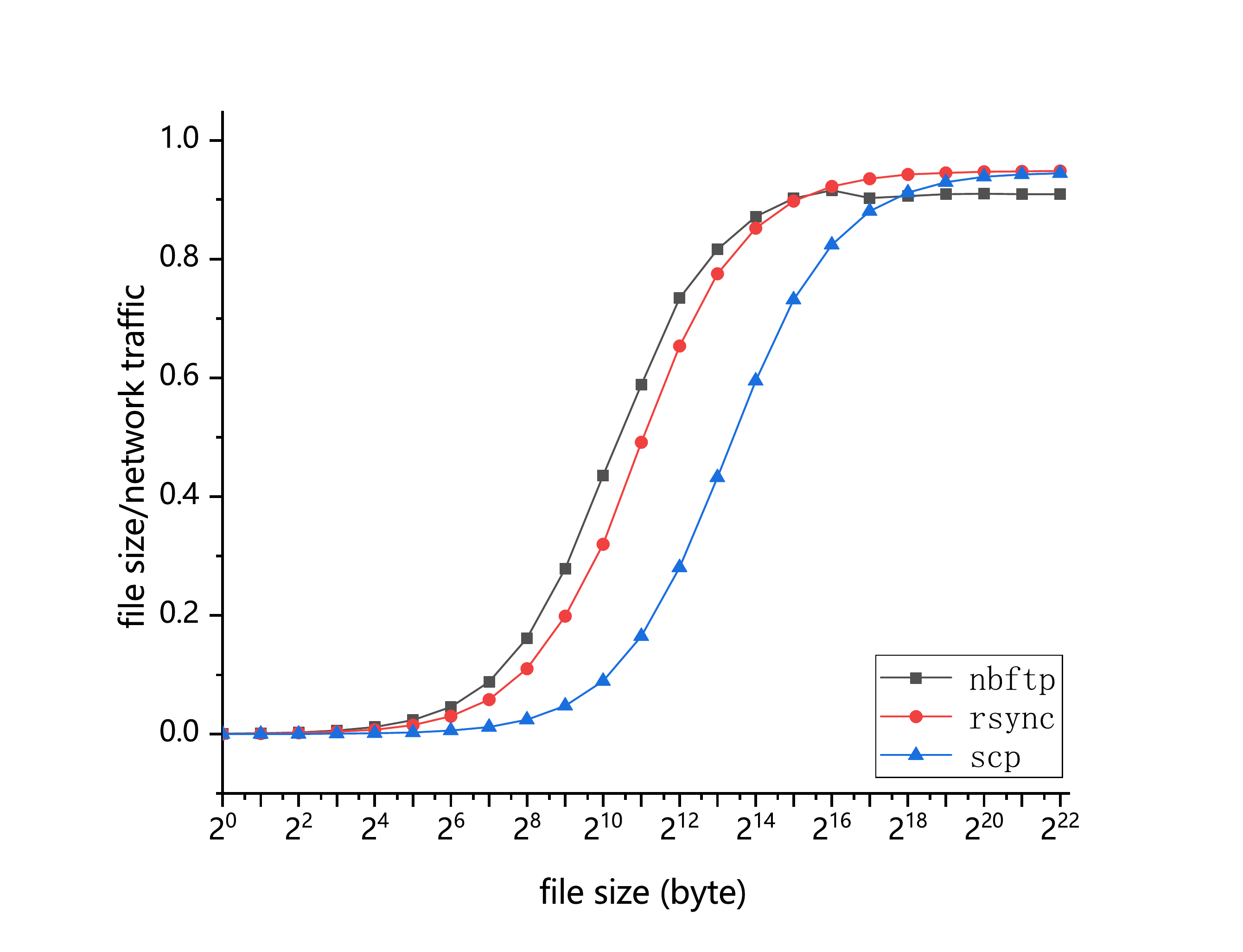}
    \subcaption{file size / network traffic}
  \end{minipage}
  \caption{The actual network traffic transmitting a single file of different sizes. The network traffic refers to the total size of the IP packets transmitted by the sender and receiver. When the size of the target file is less than $2^{15}$ bytes, NBFTP requires the least amount of network traffic. When the file size exceeds $2^{15}$ bytes, the efficiency of these three methods tends to be the same.}
\label{fig:fig2}
\end{figure}

Single file transfer is the basic function of NBFTP. We compared the network traffic of NBFTP, rsync, and scp when transferring single files of different sizes, as shown in Figure~\ref{fig:fig2}.

For the system log and site-testing data that need to be transferred periodically, a single file is usually less than 8KB. When transmitting such data, NBFTP generates less network traffic than rsnyc and scp. The reason why the actual amount of data transferred by rsync is larger is that it requires additional operations to determine which files need to be transferred. The additional data transmission required by the scp method mainly comes from the encryption of the data transmission process. When transferring larger files, the amount of network traffic required by the three methods tends to be the same.

\subsection{Directory Synchronization Overhead}
\label{sec:sec4.3}

During long-term synchronization, most files in the source directory are most likely the same as files in the remote destination directory. As mentioned earlier, rsync uses algorithms to determine the difference between the source and the target, thereby reducing network traffic. However, the frequent comparisons of the same files will consume unnecessary network traffic. The more same files, the more extra network traffic rsync consumes in the process of finding differences.

\begin{figure}[tbp]
	\centering
  \begin{minipage}{0.33\linewidth}
    \centering
    \includegraphics[width=\linewidth]{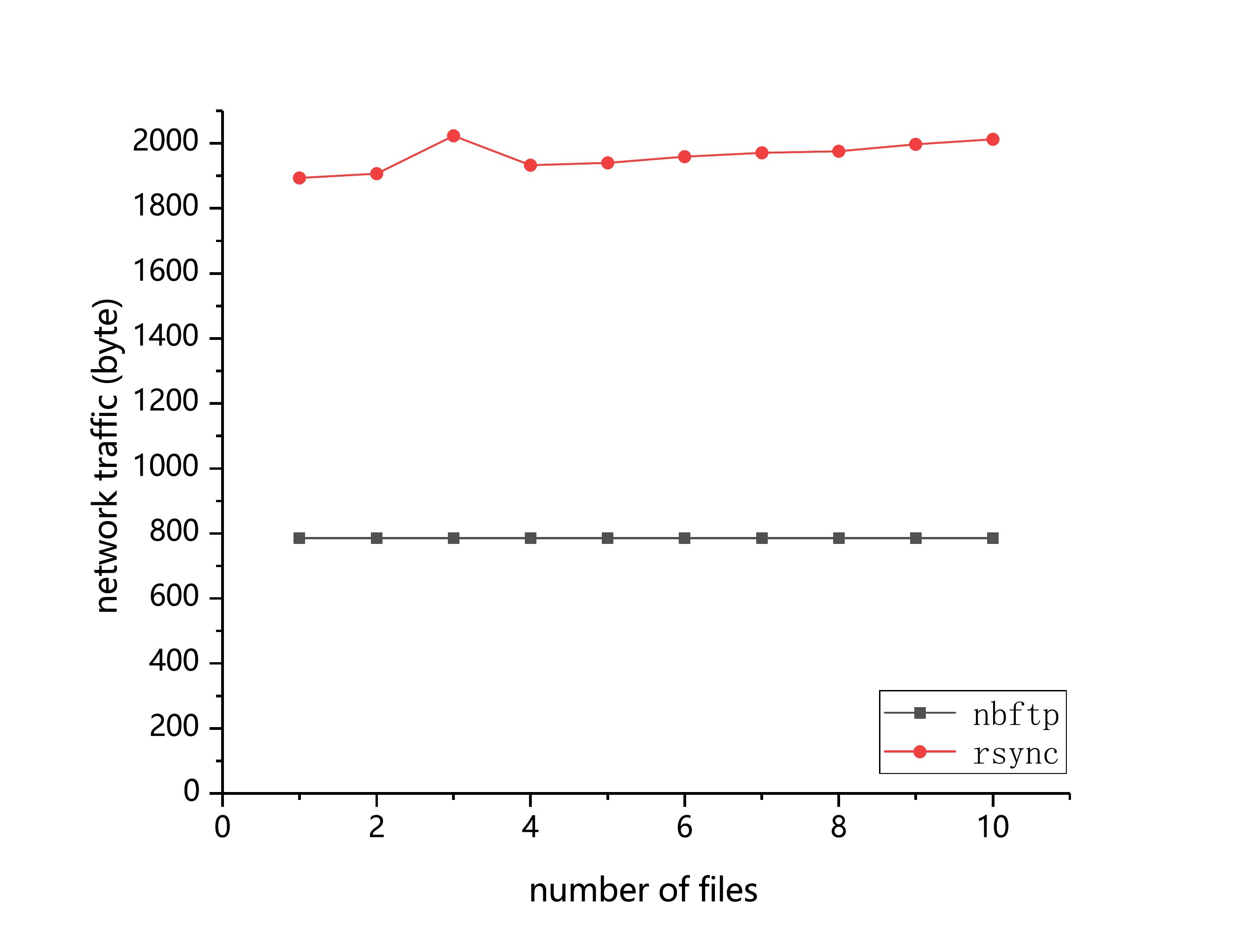}
    \subcaption{1KB}
  \end{minipage}
  \begin{minipage}{0.33\linewidth}
    \centering
    \includegraphics[width=\linewidth]{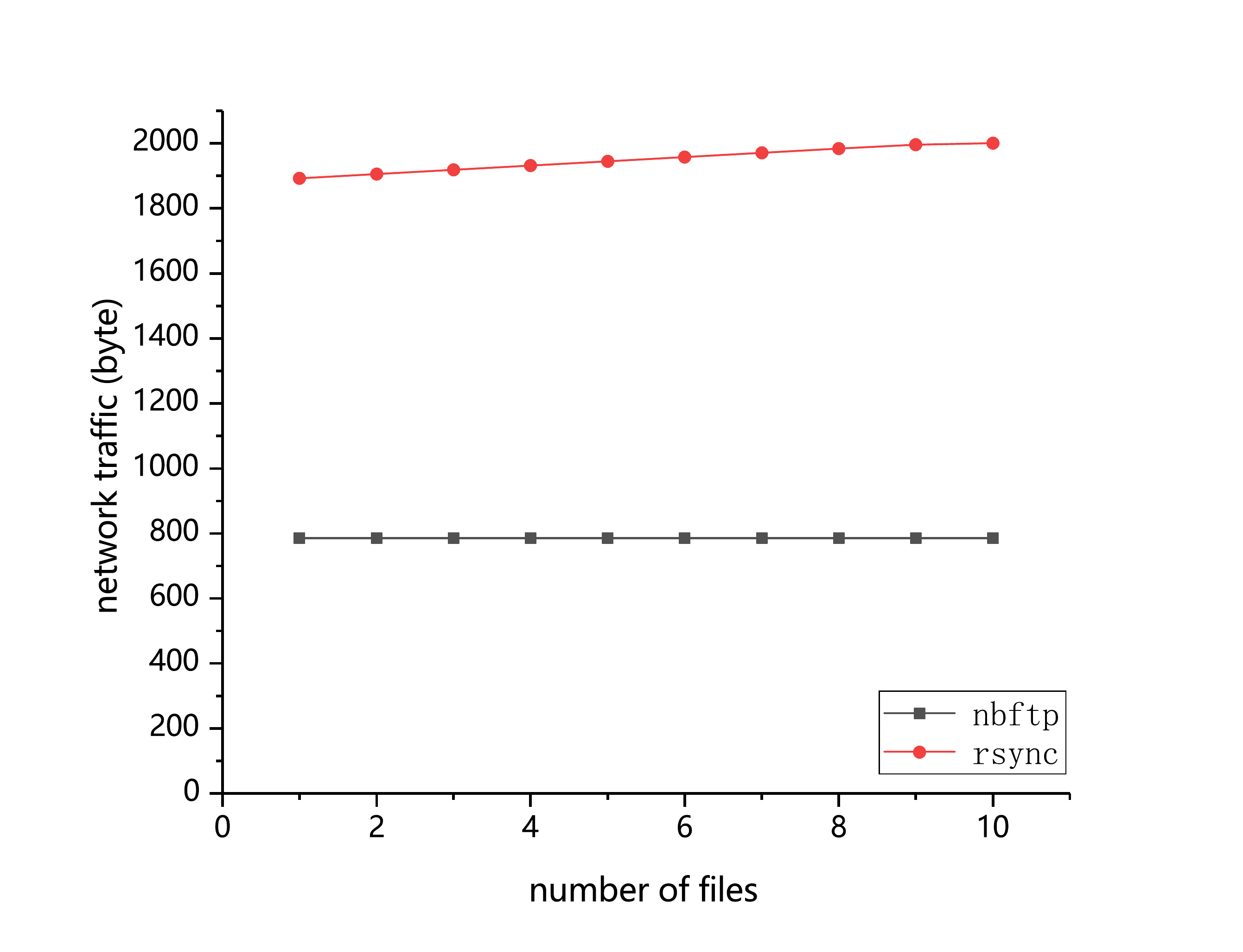}
    \subcaption{8KB}
  \end{minipage}
  \begin{minipage}{0.33\linewidth}
    \centering
    \includegraphics[width=\linewidth]{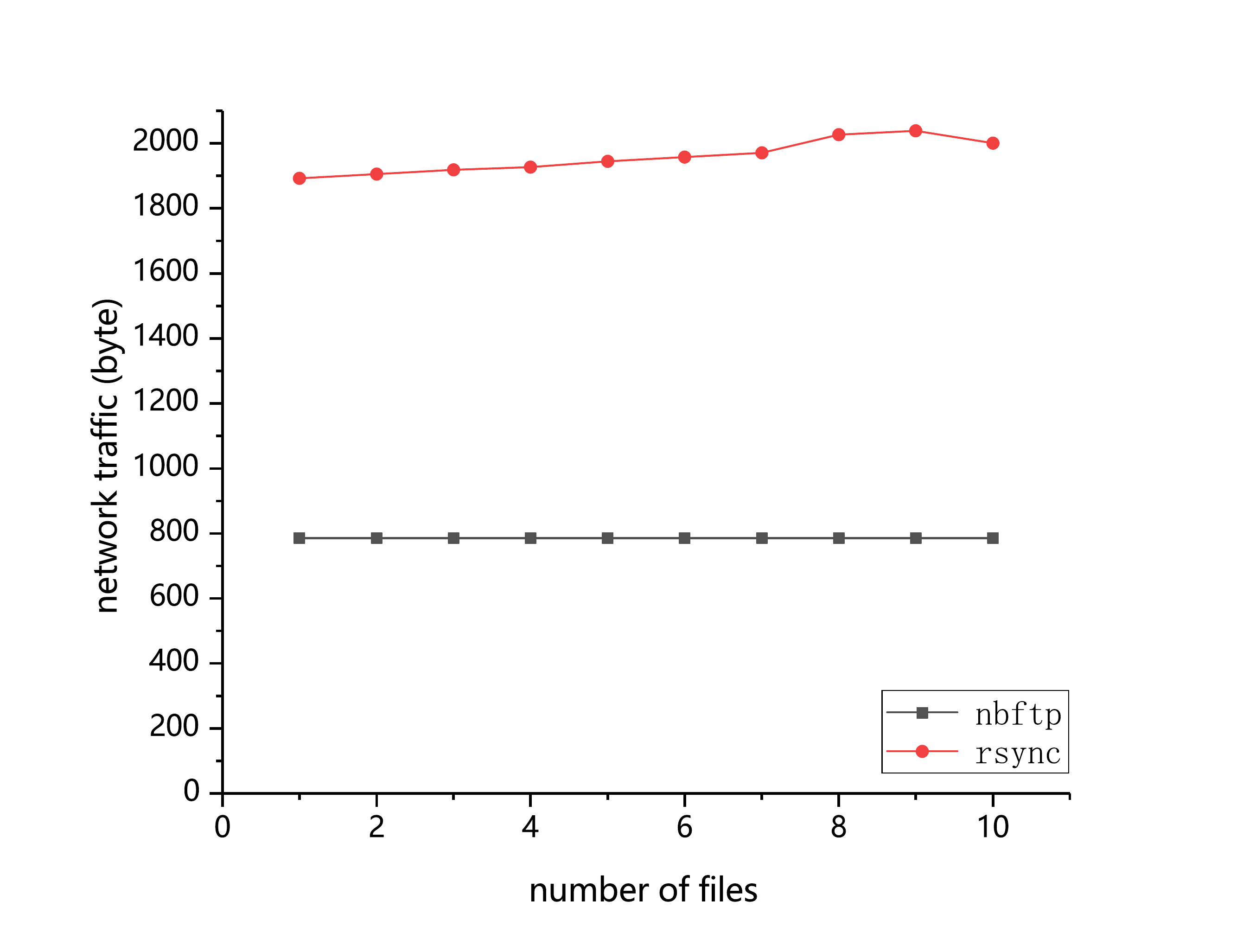}
    \subcaption{64KB}
  \end{minipage}
	\caption{The actual network traffic transmitting multiple small files that have already been transferred}
\label{fig:fig3}
\end{figure}

Figure~\ref{fig:fig3} validates our conjecture. In each test case, the file sizes are also 1KB, 8KB, and 64KB. In the beginning, the source and destination directories hold the same files, which means all files have been synchronized. Then we use NBFTP and rsync to transfer them again. The results show that the network traffic of rsync is about 2.4 times that of NBFTP on average, and increases when the size and number of synchronized files doubles, while the network traffic of NBFTP is almost unchanged. One method to reduce the overhead of using rsync is deleting the transferred files in the source directory, but it will cause potential data security issues. Overall, the directory synchronization overhead of NBFTP is less than rsync, so NBFTP is more suitable for long-term transmission between Dome A and DCC.

\subsection{Periodically Generated Small Files}

\begin{figure}[tbp]
	\centering
  \begin{minipage}{0.33\linewidth}
    \centering
    \includegraphics[width=\linewidth]{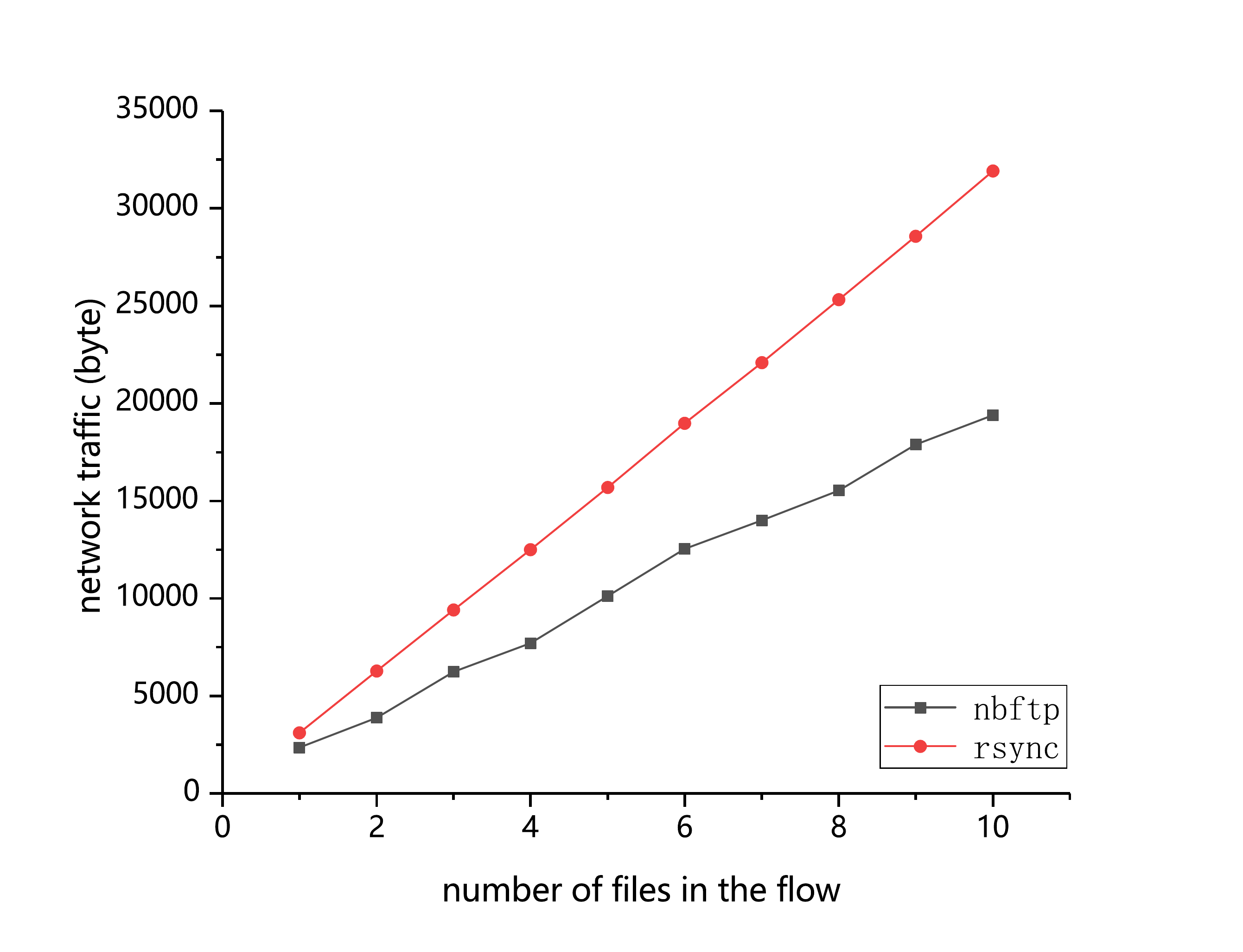}
    \subcaption{1KB}
  \end{minipage}
  \begin{minipage}{0.33\linewidth}
    \centering
    \includegraphics[width=\linewidth]{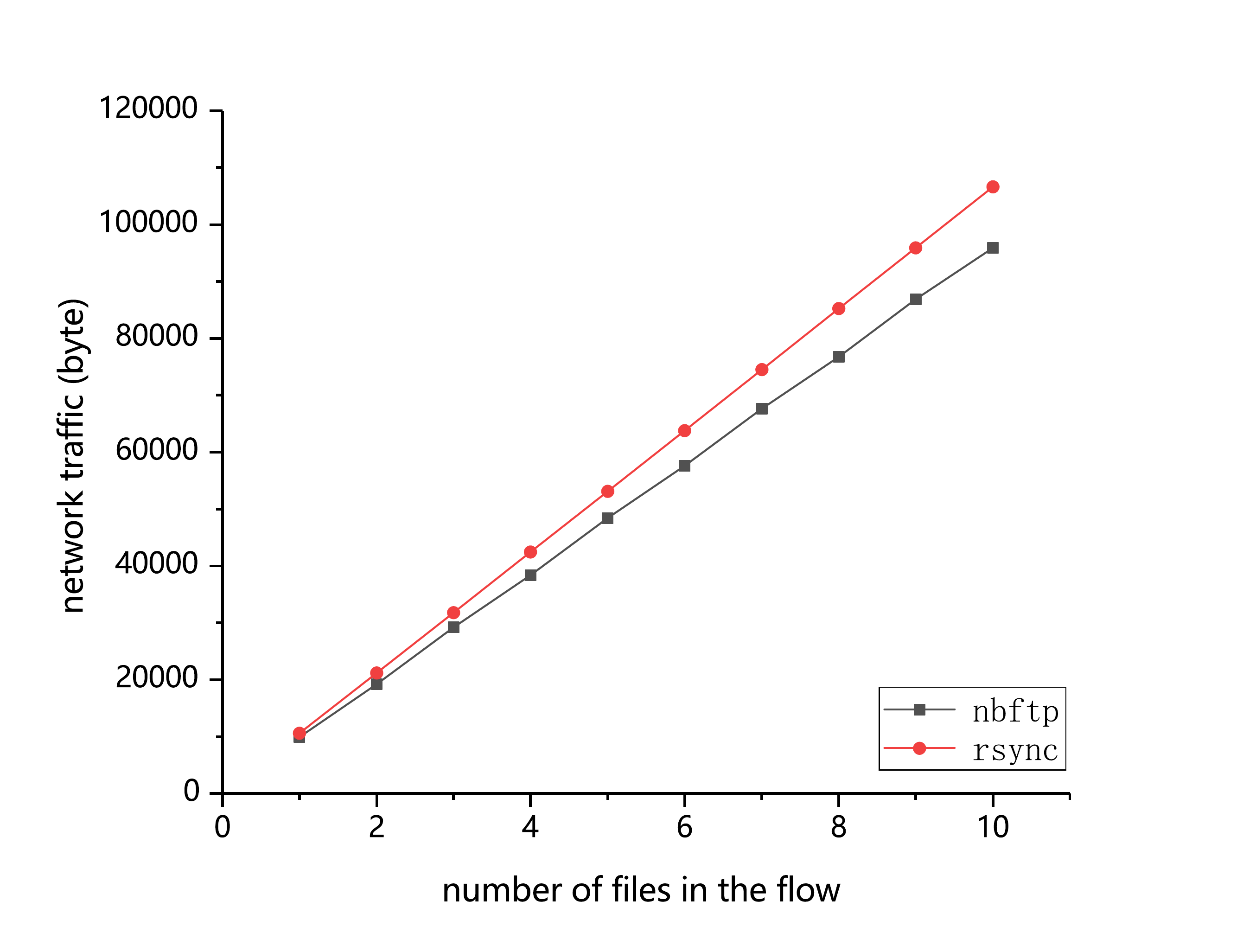}
    \subcaption{8KB}
  \end{minipage}
  \begin{minipage}{0.33\linewidth}
    \centering
    \includegraphics[width=\linewidth]{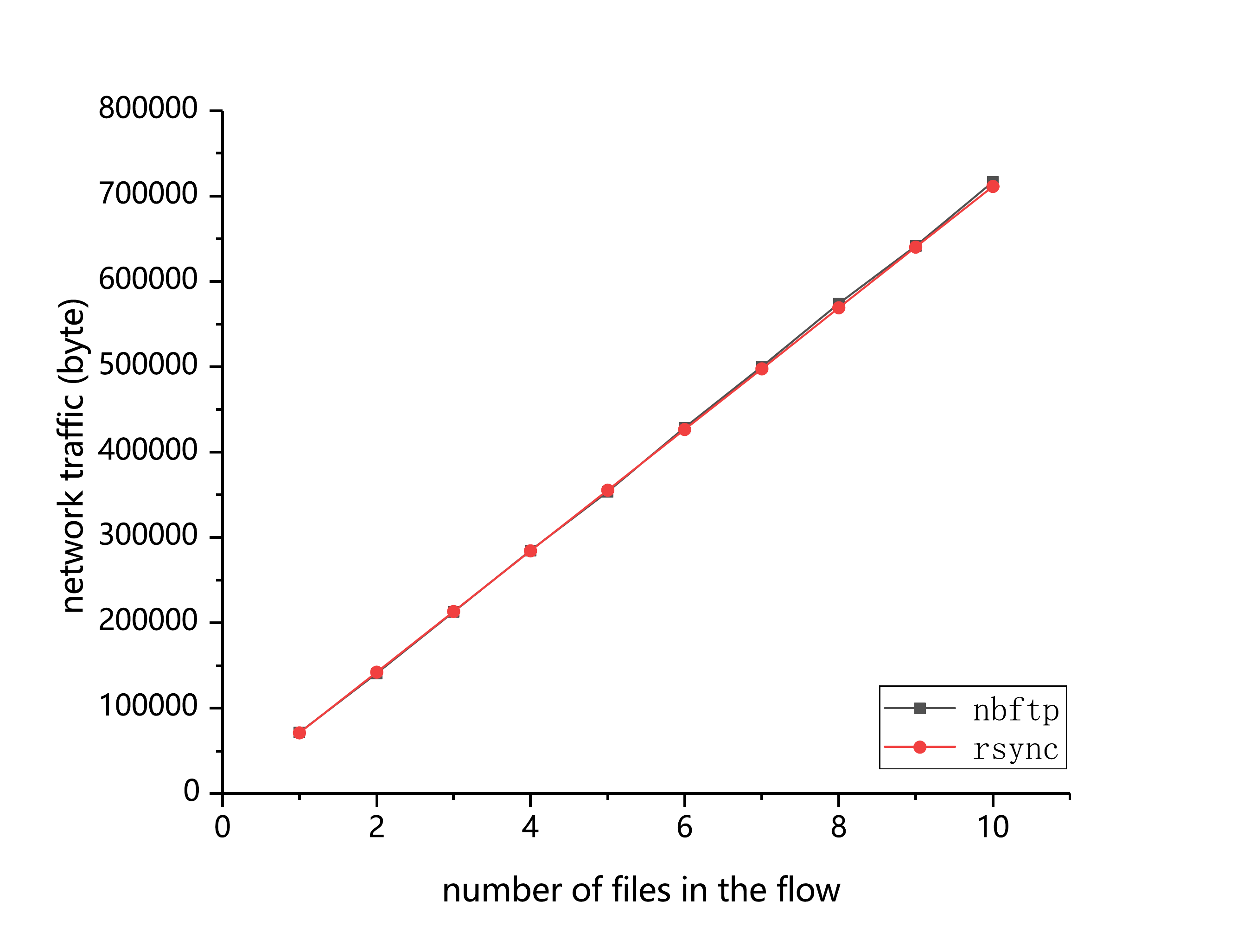}
    \subcaption{64KB}
  \end{minipage}
	\caption{The actual network traffic transmitting periodically generated small files of different sizes}
\label{fig:fig4}
\end{figure}

As mentioned earlier, the site-testing data will be transferred from Dome A periodically. In this section, we will simulate the transmission process of the actual situation. Small files are generated into the source directory at regular intervals other than immediately put into the queue to transmit. Just like the above experiments, files are of 1KB, 8KB, and 64KB respectively. Figure~\ref{fig:fig4} shows the network traffic when the number of files reached different levels.

This experiment can be regarded as a combination of the above two experiments. On the one hand, the files are generated at regular time, and each generation will trigger a transmitting operation to transfer files one by one, corresponding to the first experiment. On the other hand, the number of transferred files doubles over time, and there can be a lot of duplicates in the source and target directories, which is like the second experiment.

From Figure~\ref{fig:fig4}, it can be seen that when the sizes of files to be transferred are 1KB and 8KB, the network traffic performance of NBFTP is better than rsync due to the same reason shown in Figure~\ref{fig:fig2}. As the number of files increases, NBFTP performs better than rsync because the latter need to check more and more same files, which is shown in Figure~\ref{fig:fig3}. The network traffic performance of NBFTP improves as the number of files increases but decreases as the file size increases. In general, NBFTP can save more network traffic, because its performance is almost the same as rsync when transferring large files, and in the case of transferring smaller files, the efficiency of NBFTP is better than rsync.

\begin{figure}
  \centering
  \includegraphics[width=0.5\linewidth]{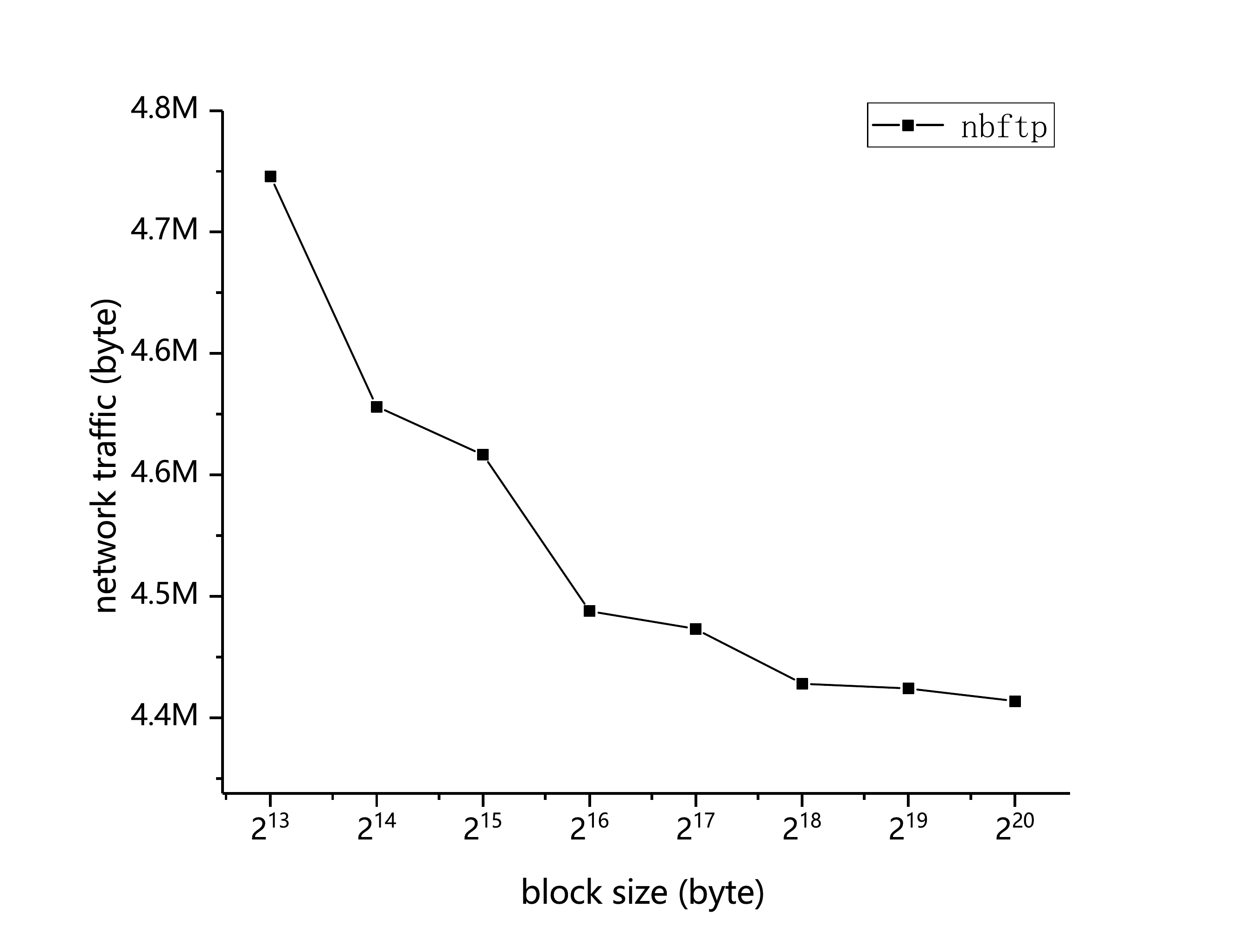}
  \caption{Impact of block size in NBFTP}
  \label{fig:fig5}
\end{figure}
`

\subsection{Impact of Block Size in NBFTP}

NBFTP divides large files into blocks for transmission. The file block size setting will affect the network traffic during transmission. When transferring files of the same size, the larger the block size, the smaller the number of blocks, the less space the header file takes up, and the less network traffic it consumes. Figure~\ref{fig:fig5} shows the network traffic of transferring 4MB files with different block sizes using NBFTP. The experimental results are as expected.

However, large blocks cause high costs for retransmissions. In order to save network traffic in the unstable Iridium network environment, NBFTP uses an adaptive block size mechanism to select the optimal block size according to the network situation. The specific introduction of the adaptive block size mechanism is in Section~\ref{sec:sec3}.

\section{CONCLUSION}

In this paper, we designed a dedicated data transfer system called NBFTP for remote astronomical observation at Dome A, Antarctic. NBFTP provides a highly stable, traffic-saving, and easy-to-use data channel by optimizing the ratio of file size to network traffic, resuming break-points, and implementing extensibility.

Iridium communication link is narrow-band and unstable, and the network traffic price is high. The retransmission mechanism of existing tools can cause extra network traffic. To this end, NBFTP controls the traffic at the block level, implements the break-point resuming function based on the adaptive block size mechanism, and effectively controls the overhead caused by retransmissions. Our experimental results show that NBFTP is superior to rsync and scp in data traffic. It consumes 60\% less network traffic than rsync when detecting the data pending to be transferred. When transferring continuously generated 1KB files, NBFTP consumes 40\% less traffic than rsync.

The reasons why NBFTP has transmission efficiency come from three levels of optimization. Firstly, NBFTP adopts a special transmission protocol that consider different file sizes to reduce extra data (see Section~\ref{sec:sec3.2}). Secondly, it uses a fine-grained data retransmission mechanism that reduces the overhead of data retransmission under unstable network conditions (see Section~\ref{sec:sec3.3}). Finally, when doing remote directory synchronization, NBFTP can avoid the overhead of full directory scanning required by rsync, based on local logs (see Section~\ref{sec:sec4.3}).

In addition, existing tools have only basic data transfer functions and cannot support automatic transaction processing. NBFTP implements an Expansion Module with triggering functions such as mail sending, data cleaning, and data archiving, which can provide more support for remote astronomical observation.

NBFTP and its data transfer protocol are specially designed for remote astronomical observations based on narrow bandwidth and unstable Iridium communication network. Table~\ref{tab:tab6} shows the comparison of related functions of NBFTP and other commonly used tools mentioned in Section~\ref{sec:sec2}.

\begin{table}
\bc
\begin{minipage}[]{\linewidth}
\caption[]{Function comparison of several data transfer tools\label{tab:tab6}}\end{minipage}
\setlength{\tabcolsep}{5pt}
\small
 \begin{tabular}{cccc}
  \hline\noalign{\smallskip}
  Function & scp & rsync & NBFTP \\
  \hline\noalign{\smallskip}
  Large file transfer & Send the entire file & Compare the difference & Split files and receive a\\
   & and only receive a & between the source and the target, & confirmation message \\
   & confirmation message & and only transfer the difference & every time a block is sent\\
  Break-point resuming & Unsupported & Supported & Supported \\
  Network traffic control & Unsupported & Unsupported & Supported \\
  Guarantee of file integrity & Unsupported & Unsupported & Supported \\
  Encryption & SSH & Unsupported & OpenSSL \\
  \noalign{\smallskip}\hline
\end{tabular}
\ec
\end{table}

A stable version of NBFTP has been serving the data transfer of KLAWS, KCLAM, KL-DIMM at Dome A. NBFTP can also be helpful for site testing and early-stage astronomical observation at other places where such narrow-band and unstable Iridium communication is the only choice. In the future work we will focus on more fine-grained network traffic control and try to expand it to support short message communication of Beidou.

\normalem
\begin{acknowledgements}
This work is supported by the Joint Research Fund in Astronomy (U1931130) under the cooperative agreement between the National Natural Science Foundation of China (NSFC) and the Chinese Academy of Sciences (CAS). HU, SHANG, and MA acknowledge support from NFSC under grant number 11873010 and 117330037, and the Operation, Maintenance and Upgrading Fund for Astronomical Telescopes, and Facility Instruments, budgeted from the Ministry of Finance of China (MOF) and administrated by the CAS.

\end{acknowledgements}

\bibliographystyle{raa}
\bibliography{ms2020-0164}

\end{document}